# Characterization of chemoelastic effects in arteries using digital volume correlation and optical coherence tomography.

Víctor A. Acosta Santamaría*, María Flechas García*, Jérôme Molimard, Stéphane Avril**

Mines Saint-Etienne, Univ Lyon, INSERM, U 1059 Sainbiose, Centre CIS, F - 42023 Saint-Etienne France

*These authors contributed equally to this manuscript.
**Corresponding author (avril@emse.fr)

**Abstract**
Understanding stress-strain relationships in arteries is important for fundamental investigations in mechanobiology. Here we demonstrate the essential role of chemoelasticity in determining the mechanical properties of arterial tissues. Stepwise stress-relaxation uniaxial tensile tests were carried out on samples of porcine thoracic aortas immersed in a hyperosmotic solution. The tissue deformations were tracked using optical coherence tomography (OCT) during the tensile tests and digital volume correlation (DVC) was used to obtain measurements of depth-resolved strains across the whole thickness of the tested aortas. The hyperosmotic solution exacerbated chemoelastic effects, and we were able to measure different manifestations of these chemoelastic effects: swelling of the media inducing a modification of its optical properties, existence of a transverse tensile strain. For the first time ever to our best knowledge, 3D strains induced by chemoelastic effects in soft tissues were quantified thanks to the OCT-DVC method. Without doubt, chemoelasticity plays an essential role in arterial mechanobiology *in vivo* and future work should focus on characterizing chemoelastic effects in arterial walls under physiological and disease conditions.

**Statement of significance:** Chemoelasticity, coupling osmotic phenomena and mechanical stresses, is essential in soft tissue mechanobiology. For the first time ever, we measure and analyze 3D strain fields induced by these chemoelastic effects thanks to the unique combination of OCT imaging and digital volume correlation.

# 1. Introduction

Arteries, which carry our blood away from the heart to all organs, have three distinct tissue layers: the *tunica intima*, the *tunica media*, and the *tunica adventitia*. In large elastic arteries such as the aorta, the media is the thickest layer with the most substantial contribution to arterial mechanics under normal physiological conditions. It is a stack of highly structured lamellas, comprising plates of densely packed elastin and collagen fibers alternating with looser collagen and elastin networks in which the fascicles of smooth muscle cells (SMCs) are embedded. The looser collagen and elastin networks are commonly surrounded by a gel-like ground substance with abundant glycosaminoglycans (GaGs) and proteoglycans [1].

Whereas the biomechanical properties of large arteries have been an intense topic of research [2], transport of water across arterial walls is a topic which received significantly less attention [3][4]. An important transmural pressure gradient (100 mmHg) exists between the intraluminal arterial blood pressure and adventitial interstitial pressure. This pressure gradient is responsible for a radial flow through the arterial wall, which depends: (i) on hemodynamic factors, including pressure and wall shear stress modulating the permeability of the endothelium [5], (ii) on the permeability of the arterial wall, which depends on the elastic network, its stretch and on the SMC tone [1], (iii) on osmotic pressures in the wall and in the surrounding media. For example, free water molecules will be retained more in hydrophilic areas of the aortic wall than in hydrophobic ones (elastin), and could contribute to the swelling of the GAG-rich areas [6][7].

As the stretch of the extracellular matrix (ECM) affects outward advection, it is evident that the interstitial fluid also plays an important role on the deformability of soft biological tissues. Its ability to move through the interstitial space and across the boundary permits reversible changes of tissue volume and fluid exchange with the environment, which in turn is affected by the osmotic activity of the tissue constituents. These fluid motions are commonly neglected and this is often taken as an argument for tissue incompressibility. In the pioneering work of Carew et al. (1968), the arterial wall was assumed as a homogeneous substance consisting in semisolid fibrillar structures embedded in a gelatinous matrix, setting the basis for nearly incompressible hyperelastic models within the framework of continuum mechanics [8]. However, hydrated biological tissues are intrinsically biphasic (solid matrix and interstitial fluid).

An incompressible hyperelastic analysis may be representative of a biphasic analysis over the short-term response [9] but beyond, the possible motion of interstitital fluid can induce significant volume variations [10,11].

Hydration may also affect significantly the mechanical behavior of tissues. For instance, elastin needs to be highly hydrated to remain elastic [12]. Wang et al. (2018), recently showed that purified elastin treated with polyethylene glycol (hence with a significant dehydration, including loss of intrafibrillar water) became very rigid and was also significantly more viscoelastic [13].

Therefore, for a comprehensive approach of arterial mechanobiology, it is necessary to model fluid flow, variations in chemical potential and osmotic pressure in addition to stress and strain in the tissue. Such coupled phenomena of fluid flow and deformation which are typical of hydrated biological tissues are well treated within the theory of chemoelasticity [14, 15], which was originally developed to address chemically driven swelling in geomechanics [16]. It has been well adopted for soft tissues such as cartilage, which contain a significant fraction of GaGs [17]. Lanir introduced a chemoelastic theory of arteries to explain residual stresses [18]. Following Lanir's pioneering work, Azeloglu et al. (2008) investigated, both numerically and experimentally, the regulating role of proteoglycans present in the arterial wall, showing that, with an inhomogeneous distribution of proteoglycans through the wall thickness, the osmotic pressure would vary across the wall thickness, resulting in an inhomogeneous swelling stress field in the solid matrix, affecting significantly the opening angle observed experimentally [19].

Despite these advanced theoretical and computational studies, chemoelasticity in arteries still requires experimental insights. Among the manifold of experimental techniques, full-field optical measurement techniques, such as digital image correlation (DIC) or digital volume correlation (DVC) [20-27], have recently been extended to the mechanical behaviour of arteries and have revealed major regional inhomogeneities of strains and stiffness [28, 29].

Additionally, and regarding the characterization of aortas, few studies have used the OCT technique to achieve full-field strain measurements [30]. Only Acosta Santamaría et al. (2017), applied a tensile test in conjunction with the OCT and DVC methods to obtain measurements of depth-resolved deformations across porcine aortic tissues immersed in a tissue clearing agent (propylene glycol - PG) [31, 32]. The obtained

OCT-DVC strain fields showed interesting inhomogeneities across the different layers of the aorta.

In the work reported here, similar experiments were repeated and analyzed thoroughly, showing that PG induces incidentally major chemoelastic effects on arteries which permit to observe fluid flow and their induced strains in arteries undergoing uniaxial tension [33]. The objective of the present paper is then to characterize chemoelastic strain fields in arteries using OCT-DVC.

## 2. Materials and methods

### 2.1. Sample preparation

Four descending thoracic aortas were ordered from the Veterinary Campus of the University of Lyon (VetAgro Sup, Marcy l'Etoile). Youna castrated pigs, aged between 5 and 6 months and weighing between 80 and 90 kg were sacrificed in accordance with the recommendations of the VetAgro Sup Ethics Committee (C2EA No. 18), and in accordance with the regulations on animal experimentation - Directive 2010/63 / EU. Aortas were excised and stored at −20°C. The postmortem autolysis was minimized by an immediate freezing of the tissue and by performing the tests soon after the specimens were thawed.

After thawing, and for all the experimental protocols implemented, the samples were immersed in an osmotically active solution one hour before and during the tests (see figure 1a). Propylene glycol (PG) and phosphate buffered saline (PBS) were mixed to prepare an osmotically active solution improving the optical properties of the tissue. Accordingly, the obtained penetration depth and image contrast enabled measurements across the whole thickness of the aortic wall [31]. The PG, as an optical clearing agent, was previously used by our group. Details may be found in our previous publication [31].

All the aortic samples were obtained in a region between intercostal branches (see figure 1a). The length was aligned with the circumferential direction of the tissue (Y-Axis), and the width with the longitudinal direction (X-Axis), whereas the Z-Axis defined the thickness (see figure 1a).

In total, 14 aortic samples were prepared for different purposes:
- one sample (10 x 14 mm), named sample A1, was used to perform the OCT-DVC calibration using a rigid body motion test.

- six samples (10 x 20 mm), named samples B1, B2, …, B6, were used to evaluate the optical scattering properties of the aortic tissue immersed in different osmotically active solution concentrations.
- three rectangular samples (10 x 58 mm), named C1, C2 and C3, were immersed in an 80% PG solution and tested in uniaxial tension (stepwise stress-relaxation tests). Strain fields were measured during these tests using the OCT-DVC technique described further.
- two samples (10 x 58 mm), named D1 and D2, were immersed in a 60% PG solution and tested in stepwise uniaxial tension for monitoring the evolution of the OCT signal. Strain fields were measured during these tests using the OCT-DVC technique described further.
- two samples (10 x 58 mm), named E1 and E2, were immersed in a 100% PBS solution and tested in stepwise uniaxial tension for monitoring the evolution of the OCT signal without using the OCT-DVC method.

For stepwise uniaxial tensile tests (samples C1, C2, C3, D1, D2, E1 and E2), the wall thickness was measured before and after the immersed condition and at the end of the experimental test (Table 1). Two methods were applied. For each specimen, the thickness was measured using a digital caliper while holding the specimen between two plates (5 different locations). Considering the OCT acquisitions (A-Scan's), additional measurements were obtained directly in the middle zone of the field of view (5 different locations). The reported difference between the two techniques was less than 5%.

## 2.2. OCT acquisitions

A preliminary calibration of the OCT-DVC method was achieved on sample A1. A rigid body translation test was established to determine the optimal voxel size of OCT acquisitions. Linear translations were applied in the X-Axis (0, 20, 40 and 60 µm), and considering 5, 6, 7, 8 and 9 µm pixel sizes on the longitudinal-circumferential X-Y plane (OCT B-Scans). Other pixel sizes were explored in our previous publication (3, 4, 5 and 6 µm) [31]. For the Z-Axis the pixel size was 1.42 µm (OCT A-Scan). The field of view (FOV) was 2 x 3 x 1.48 mm (X, Y and Z-Axis, respectively).

Additionally, and to determine the incidence of the optical clearing agent on the optical properties of the tissue, several PG concentrations were considered on samples B1 to B6. The goal was to identify the minimum and necessary PG concentration to allow sufficient contrast for an entire through-thickness OCT acquisition. The concentrations varied between 30% and 80% v/v PG/PBS: 30/70 for sample B1, 40/60 for sample B2, 50/50 for sample B3, 60/40 for sample B4, 70/30 for sample B5 and 80/20 for sample B6. OCT acquisitions were performed with 6 µm pixel size on the X-Y plane and 2.45 µm pixel size in the Z-Axis. A FOV of 2 x 2 x 2.8 mm was defined.

Subsequently, the OCT signal intensity was acquired during stepwise uniaxial tensile tests [31] carried out on samples C1, C2 and C3 immersed in the 80% PG solution. A stepwise stress-relaxation ramp was applied and OCT volumetric images were acquired for each load-step at the end of the stress-relaxation phases. The OCT parameters were: a voxel size of 5 x 5 x 2.45 µm and a FOV of 2 x 4 x 2.51 mm (X, Y and Z-Axis, respectively), see figure 1b. In comparison to the rigid body translation test, and to ensure that the zone of interest deformed without significant rigid body motion, the FOV was increased of one millimeter in the Y-Axis. We used a Thorlabs OCT- TEL220C1 OCT system. The system features were: a center wavelength of 1300 nm, lateral resolution 7 µm, focal length 18 mm, maximum sensitivity range 111 dB (at 5.5 kHz). During the experiments, the OCT illumination tube was in full-contact with the osmotic solution and the lens was focused on the outer surface of the intima layer (corresponding to OCT B-scans), see figures 1a - 1b. The OCT acquisition datasets were saved in TIFF format. The TIFF virtual stack was imported using the ImageJ® software. The data were rescaled and converted to 8 bits for intensity levels digitalization. Finally, the data were exported as RAW images. With the RAW data, and applying the DVC method, displacement fields were measured.

Then, a similar stepwise uniaxial tensile test was carried out on samples D1 and D2 immersed in the 60% PG solution. Finally, the stepwise uniaxial tensile tests was carried out on samples E1 and E2 immersed in the 100% PBS solution. No OCT volumetric acquisitions were acquired for these samples.

**2.3. Uniaxial tensile test with stepwise stress-relaxation**

Before carrying out the stepwise uniaxial tensile tests, on samples C1, C2, C3, D1, D2, E1 and E2, two preconditioning cycles were applied [34-37]. Both cycles considered a 1.15 stretch of the sample and return to the unloaded length. The initial distance between clamps was 35.1 mm (see figure 1a).

After these two cycles, a 1.15 stretch was applied again (displacement increment of 5.2 mm). The displacement was maintained during 38 min, 30 min to define the equilibrium and 8 min to obtain the corresponding OCT acquisition. During these 38 minutes, the sample underwent relaxation and eventually stabilized as the final relaxation rate was below 100 Pa/min [38].

After preconditioning, a stepwise stress-relaxation ramp was applied with nine controlled displacements and relaxation phases between increments of 38 min. The displacement ramp was defined in the circumferential direction of the tissue by increments of 0.6 mm (Y-Axis, corresponding to OCT B-scans), see figures 1b - 1c. In order to maintain the center of the sample in the same position, equal displacements were applied in two opposite directions simultaneously (0.3 mm on each side), see Fig 1c. A final stretch of 1.28 was approximately reached. The load was monitored with a submersible load cell of 22 N (rated output +/-1.57 mV/V) conditioned with a Futek IPM650 panel mount display (input range up to +/-500 mV/V).

### 2.4. Strain measurements by Digital Volume Correlation (DVC)

3D displacement fields were measured using DVC with a local correlation algorithm (LA-DVC) [31, 39-41]. Briefly, the reference configuration ($f$) can be represented by the gray-level function $f(x, y, z)$, and the deformed state ($g$) as $g(x + u, y + v, z + w)$, where $(x, y, z)$ represents the coordinates and $(u, v, w)$ the offset in each direction. Considering $(u, v, w)$ as the displacement mapping, the continuity of the gray-level can be assumed as $f(x, y, z) = g(x + u, y + v, z + w)$ [41, 42]. The DVC method was applied using the DaVis® (LaVision) software. The sub-volume discretization and the multi-pass approach in DaVis® were used to achieve the maximum cross-correlation coefficient (considering the gray-level distributions) [40, 43-45].

The implemented correlation parameters were the same as in our previous work [31], but we added a uniform filter and a 2 x 2 x 2 binning volume. For the rigid body translation test on sample A1, the DVC method was performed at 3 different steps and permitted to reconstruct the displacement fields between the reference configuration and each of these translated configurations. In the stepwise stress-relaxation uniaxial tensile tests, the DVC method was performed at 9 different steps and permitted to reconstruct the displacement fields between the reference configuration (defined after preconditioning) and each of these deformed configurations.

Before performing DVC, four specific regions of interest were defined in the OCT volume images: the whole aortic wall (global), the intima region (layer I), the media region (layer M) and the adventitia region (layer A). After measuring the 3D displacement fields in each region, they were approximated by tricubic functions and the Green–Lagrange strain components were derived [46-48]. All the postprocessing approach was implemented in MatLab®.

### 2.5. Identification of the short-term stiffness

Our stepwise loading protocol was decomposed into nine steps, and each step consisted in a displacement ramp (phase 1) followed by a relaxation/swelling phase of about 30 min (phase 2). In Appendix 1 (supplemental materials), a biphasic model accounting for the relaxation/swelling effects is proposed. According to [9], it is assumed that fluid motion is negligible during the displacement ramp (phase 1) which represents the short-term behavior during which the tissue remains almost incompressible. Then, for each of the nine loading steps performed in our experiment, the following equation could be derived after Eq (A11) (supplemental materials) to estimate the short-term stiffness parameter:

$$\bar{\mu}^S = \frac{\Delta\sigma}{\Delta\lambda} \frac{\bar{J}^{2/3}}{1+\frac{\bar{\lambda}_x^2+\bar{\lambda}_y^2}{\bar{\lambda}^2}} \qquad (1)$$

where $\Delta\sigma$ and $\Delta\lambda$ are respectively the difference of stress and stretch between the beginning and the end of phase 1, whereas $\bar{\lambda}$, $\bar{\lambda}_x^2$, $\bar{\lambda}_z^2$ and $\bar{J}$ are respectively the axial stretch, the stretch along the $x$ direction, the stretch along the $z$ direction and the change of volume taken at the beginning of phase 1 of each step.

### 3. Results

### 3.1 OCT acquisitions

The OCT-DVC method was preliminarily optimized in terms of spatial resolution and contrast acquisition. For the spatial resolution, the 5 µm pixel size provided the smallest DVC errors, yielding displacements of 19.2 ± 3.12 µm, 39.9 ± 3.78 µm and 60.4 ± 4.64 µm, for applied translations of 20 µm, 40 µm and 60 µm, respectively. For the contrast, acquisitions across the entire aortic wall were available between 60% and 70% PG concentration. Moreover, and for 80% PG, it is interesting to notice that the scattering is reduced significantly at the middle of the thickness (Figure 2). This region corresponds to the media layer of the aorta. In soft tissues, the scattering origin is attributed to the long collagen and elastin fibers and the extracellular medium which have a different refractive index than interstitial and intracellular water. After dehydration of the media the refractive index becomes nearly uniform.

The stepwise uniaxial tensile test on sample D1 and D2, immersed in a 60% PG solution permitted to assess how the OCT signal varies with the applied deformation. Figure 3 shows the OCT acquisitions for four of the nine load-steps on sample D1. The initially dark region at the middle of the thickness gradually faded away as the tissue recovered its scattering properties when the stretch increased (see figure 3). This was induced by tissue rehydration and swelling. In this context, a concentration of 60% PG does not achieve the optimal contrast and depth capability to measure 3D displacement fields across the whole aortic wall. Consequently, displacement fields could only be measured for samples C1, C2 and C3 which were immersed in an 80/20 (v/v - PG/PBS) solution.

### 3.2 Response curves

When carrying out the stepwise tensile test, temporal evolutions of the force showed a significant relaxation for samples immersed in PG (see figure 4a), as suggested by Eq (A12) (supplemental materials). Relaxation was marginal for samples E1 and E2, immersed in PBS, but it tended to be amplified at larger strains for these two samples. For samples C1, C2 and C3, immersed in 80% PG, the reported force values were averagely twice larger than the ones for samples D1 and D2, immersed in 60% PG, which again were significantly larger than the ones for samples E1 and E2, immersed in PBS. This may be explained by the effect of dehydration which stiffens the tissue. Moreover, long-term stress-strain curves were plotted, using the force values reached at the end of relaxation. The behavior shows only a marginal increase of the long-term stiffness when the PG concentration is increased (see figure 4b).

### 3.3 Strain fields

Thanks to the OCT-DVC method, the Green-Lagrange strain fields $E_{ij}$ could be measured in each sample throughout the different loading stages. The reference configuration was defined as the configuration reached after preconditioning.

In figure 5, we show the distributions of $E_{xx}$, $E_{yy}$, $E_{zz}$, $E_{xy}$, $E_{xz}$ and $E_{yz}$ across the transverse plane normal to the loading axis located at the middle of the region of interest. As expected in a tensile test, $E_{xx}$ and $E_{yy}$ had rather uniform distribution, with

$E_{yy}$ being positive (tensile strain) and $E_{xx}$ being negative (transverse strain). However, $E_{zz}$ always showed a very specific pattern, with negative strains at the top and bottom of the region of interest (intima and media layers, respectively), and positive strains at the middle of the thickness (media layer). The positive $E_{zz}$ was counterintuitive as a tensile test on an incompressible material would induce negative transverse strains along the Z-Axis. However, the positive $E_{zz}$ agrees with the model of Eqs (A12) to (A19) (supplemental materials), with fluid flowing back into the tissue when the strain increases, inducing a swelling effect at the middle of the thickness (the media layer).

### 3.4 Evolution of average strains

Figure 6 reports the force variations against the variations of the Green Lagrange strains. Average values of the strains are obtained across the whole zone of interest. It appears that $E_{zz}$ does not vary and remains almost zero when the stress increases, $E_{yy}$ reaches values between 0.12 and 0.14, and $E_{xx}$ reaches values between -0.02 and -0.04. The change of volume is defined by:

$$J = \sqrt{(2E_{xx} + 1)(2E_{yy} + 1)(2E_{zz} + 1)}$$

$J$ reaches values up to 1.1, corresponding to a global increase of volume of 10%, which may be explained by the swelling effect.

In Fig 7, we report the average engineering strain value by layers. Although $E_{yy}$ and $E_{xx}$ had nearly the same values in each of the three layers (see figures 7a and 7b), $E_{zz}$ showed a very specific response with negative strains between -0.06 and -0.08 in the intima and in the adventitia, and positive strains reaching about 0.02 in the media layer (see figure 7c). This means that the volume increase was mainly concentrated in the media, reaching 12%, whereas it was less than 5% in the intima and in the adventitia.

### 3.5 Short-term stiffness parameter

Eq (1) was used to estimate the short-term stiffness parameter for each sample and for each of the 9 loading steps. The mean value (out of 9) and the standard deviation for each sample are shown as a bar graph in Fig 8. It can be noticed that the short-term stiffness parameter is almost doubled between 60% PG and 80% PG whereas there is a factor 5 to 6 between the short-term stiffness parameter in 80% PG and the short-term stiffness parameter in PBS. This may be explained by the effect of hydration and by the effect of matrix porosity on $\mu^s$.

The standard deviations for E1 and E2 (immersed in PBS, with 0% PG) are relatively high, more than 30% of the mean value. This is explained by the significant increase of $\bar{\mu}^s$ at each loading step for these samples. Indeed, it is well known that the stress-strain behavior of arteries tends to become stiffer when the strain increases [49] due to gradual collagen recruitment. However, this stiffening behavior was lost in the samples immersed in 60% and 80% PG.

The increase of stiffness with strain for samples E1 and E2 can be observed in Fig 4a where the force jump at phase 1 tends to become larger for the final loading steps. This seems also to correlate with a more pronounced relaxation effect in these samples. Whereas the elastic behavior is stiffer when the samples are immersed in PG, our data reported for the tests in PBS are in good agreement with tests performed by other authors on porcine thoracic aortas. For instance, although Peña et al [51] and Schroeder et al [52] identified a different model than ours, our $\Delta\sigma/\Delta\lambda$ ratios are in very good agreement with their data over the 1.15 to 1.26 range of stretches used in our study.

## 4. Discussion

In this study, we carried out stepwise uniaxial tensile tests on arteries immersed in a hyperosmotic solution exacerbating chemoelastic effects, and we were able to measure different manifestations of these chemoelastic effects: tensile stress relaxation, swelling of the media inducing a modification of its optical properties, existence of a transverse tensile strain. For the first time ever to our best knowledge, 3D strains induced by these chemoelastic effects in soft tissues were quantified thanks to the OCT-DVC method.

It should be noticed first that immersion in a hyperosmotic solution (80% PG) led to a significant dehydration of the tissue, permitting to image its reflectance across the whole arterial thickness with the OCT method and thus enabling the 3D strain measurements using DVC. Dehydration is a commonplace technique for imaging soft tissues. The highly scattering nature of nontransparent human tissue limits the imaging depth of optical coherence tomography (OCT) to 1–2 mm. When longer wavelengths are used, the penetration depth can be improved; however, the imaging contrast is decreased, largely because of reduced backscattering at the microscopic scale and reduced refractive heterogeneity of the macroscopic scale. To enhance the penetration depth and imaging contrast in OCT imaging, the use of PG is rather efficient [33].

A hyperosmotic agent such as PG can draw fluid out of the tissue by creating an osmotic pressure difference between the immersion solution and the arterial fluids. The osmotic pressure inside the arterial tissue is partly regulated by GAGs and proteoglycans, which control hydration and water homeostasis in arteries and which majorly reside in the media [53]. The osmotic pressure of inflated GAGs induces tensile loading of the fibrous network (elastin, collagen), which in turn entangles and restrains the GAGs within the ECM, as well as provide shear stiffness to the ECM and help maintain ECM organization [54]. Out of a hyperosmotic agent, GAGs permit to maintain an osmotic pressure larger inside the tissue than outside, keeping water inside the tissue.

Following [9], we assumed that fluid motion was negligible during the short-term tensile response and that the response could be fitted by an incompressible Neo-Hookean model during that phase. The fitted short-term stiffness parameter showed a significant increase when the tissue was immersed in PG and the stiffening effect was larger for 80% PG than for 60% PG. Yeh and Hirshburg [56] submitted that optical clearing agents not only dehydrate the tissues but they also screen noncovalent bonding forces at the micro- and nano-scales, which may contribute to the observed stiffening effect and other possible thermomechanical manifestations [57]. Note however that the effect is reversible upon rehydration in PBS due to the ability of non-covalent forces to be recovered [56].

It is worth mentioning that the effect of dehydration was also examined by other authors in porcine aortic valve cusps [55]. Shear testing was performed on physiologically hydrated, superhydrated, and dehydrated cusps. The effect of altered hydration on shear properties was significant and followed trends which are consistent with the ones reported in our study.

A higher PG concentration improved the optical scattering properties and depth capability (see figure 2). PG dehydrated the tissue extracting the interstitial and intracellular water content. The thickness decreased by 23.10% and 23.65% after the immersion in the hyperosmotic solution, 80% PG and 60% PG respectively (see Tab 1). For Wang et al. (2018), the thickness of the aortic tissue was more sensitive to water loss and decreased to almost half of the original value using a higher solute concentration (polyethylene glycol - PEG) [13].

In this work, we only measured the steady-state strains with DVC, when the tissue had reached a chemoelastic equilibrium. The steady state situation was reached after a transient period during which fluid motions induced a relaxation response of the tissue, similar to viscoelasticity. We noticed that the duration of relaxation was increased with higher PG concentration, see figure 4. Shahmirzadi et al. (2013) also showed that relaxation effects in a dehydrated tissue (immersed in PolyEthylene Glycol) were slower than in a hydrated tissue (native - bovine thoracic aortas) [59]. Wang et al. (2018 a - b) showed that relaxation was increased after removing extrafibrillar water from isolated elastin extracted from porcine thoracic aortas [13]. Similarly in isolated elastin, a glucose treatment was also shown to increase relaxation [60]. This can be explained grossly using the biphasic model proposed in Appendix 1 (supplemental materials), showing that the tissue would be subject to an exponential decay (relaxation effect) after tensile loading. The exponential decay constant would be proportional to the stiffness $\mu^s$ of the solid matrix, the latter being significantly increased after dehydration. No one before us had measured the steady-state 3D strain fields related to osmotic effects in soft tissues. The full-field measurements performed on the aortic wall immersed in PG highlight the complexity of chemoelastically induced strains. The biomechanical behavior of arterial walls is known to be complex and anisotropic due to its fibrous nature (elastin and collagen) and due to its layered structure (intima, media & adventitia). Chemoelastic effects bring another layer to this complexity. This results in very peculiar Poisson's effects (transverse strains). An original aspect of the local mechanical behavior revealed by OCT-DVC method was the heterogeneity of transverse strains across the thickness. The intima and the adventitia showed the major effects on the mechanical behavior with corresponding transverse strains of 4.58% and 3.80% (for the maximum tensile strain of about 13%). Conversely, the media layer showed a tensile transverse strain (swelling effect). Such effects induced by chemoelasticity were already reported for thin membranes [11].

These results permit a better understanding of soft tissue biomechanics. However, there are a number of limitations that need to be reported. One of them is the number of samples. Another important one is related to the very specific testing conditions:

- use of a non physiological substance to induce artificially the osmotic effects,
- use of stepwise uniaxial tensile tests where the sample is cut out of the artery may generate new surfaces out of which the fluid can flow and this may increase water flux,

- use of frozen and thawed tissues which may have non physiological hydraulic permeability properties, as for instance freezing and thawing may cause the destruction of the cell membranes.

Harrison and Massaro [58] measured the water flux through porcine aortic tissue due to a hydrostatic pressure gradient and found it is very small, indicating that the volume changes *in vivo* would be much smaller or slower.

Therefore, there remains a very important gap towards understanding and characterizing osmotic effects *in vivo*. It is known that water filtration across the aortic wall is essential the normal physiological function of the circulation and its alteration may have important mechanobiological consequences [5]. Moreover, swelling of the arterial wall may even become detrimental as alterations in the hydro-seal might also play a role in the dissection of arterial walls when the distribution of pressures across the wall is not properly controlled [7]. Although the water permeability of the intact arterial wall has been extensively characterized [58, 61-64], the precise location and biochemical composition of the water permeation barrier, which are fundamental properties of the arterial wall, are still partially unknown and should be investigated further to better account for the essential role of chemoelasticity in arterial mechanobiology.

## 6. Conflict of interest
The authors declare they have no conflict of interest related to this work.

## 7. Acknowledgements
The authors are grateful to the European Research Council for ERC grant Biolochanics, grant number 647067.

List of Tables

Table 1. Thickness measurement and its variation depending on the osmotically active solution concentration.

| v/v PG/PBS | Thickness Measurement Conditions (mm) | | | |
| --- | --- | --- | --- | --- |
| | Before (Immersed Condition) | After (Immersed Condition) | After (Experimental Protocol) | Variation (Before / After - Immersed Condition) |
| 80/20 | 1.87 ± 0.28 | 1.44 ± 0.30 | 1.17 ± 0.40 | 23.10 % |
| 60/40 | 1.65 ± 0.30 | 1.26 ± 0.22 | 1.14 ± 0.11 | 23.65 % |
| 0/100 | 1.65 ± 0.30 | 1.44 ± 0.22 | 1.32 ± 0.19 | 12.72 % |

List of figures

1. Preparation and testing of rectangular thoracic aortic samples. (a) Collection and immersion of the samples. (b) Field of view defined for the OCT volume. (c) Experimental setup: 1- OCT system. 2- Submersible load cell. 3- Clamps. 4- Extension arms. 5- Linear pulling translation stages. 6- Immersion bath. 7- Fixed structural bases.

2. Effect of PG concentration onto the optical properties of the aortic tissue. Gray-level representation of the OCT signal intensity for six PG concentrations (different osmotically active solutions v/v - PG/PBS). Picture of the arterial samples immersed in different PG concentrations showing the evolution of the optical properties.

3. Effect of tensile strains on the reflectance of sample D1. From left to right, the stretch increases from 1.15 to 1.28, showing an increase of the reflectance signal in the middle of the artery (media layer).

4. Experimental data reported under a stress-relaxation uniaxial tensile test. a) Force-relaxation curves obtained for aortic samples immersed in osmotically active solutions with different PG concentrations (samples C1, C2 and C3 in 80% PG, samples D1 and D2 in 60% PG, and samples E1 and E2 in 0% PG). b) Tensile stress-stretch curves measured for each test. The stress values reported in this figure are derived from the force values measured at the end of the relaxation phase (phase 2) for each loading step. The stretch values reported in this figure are derived after dividing the current length of the sample (measured at the end of the relaxation phase – phase 2 – for each loading step) by its initial length (35.1 mm between the clamps at the beginning of each test).

5. Green-Lagrange strain fields measured for the maximum applied load during the tensile tests on samples C1, C2 and C3, immersed in 80% PG and 20% PBS. The zone of interest corresponds to the middle zone of the FOV defined in Figure 1.

6. Evolution of the average Green-Lagrange strains with respect to the applied force for samples C1, C2 and C3. Curves with standard deviation bars.

7. Evolution of the average Green-Lagrange strains with respect to the applied force for different regions of interest (global, intima, media and adventitia) [(a) $E_{xx}$, (b) $E_{yy}$, and (c) $E_{zz}$].

8. Identified short-term stiffness parameter (mean value and standard deviation bars) for each of the tested samples.

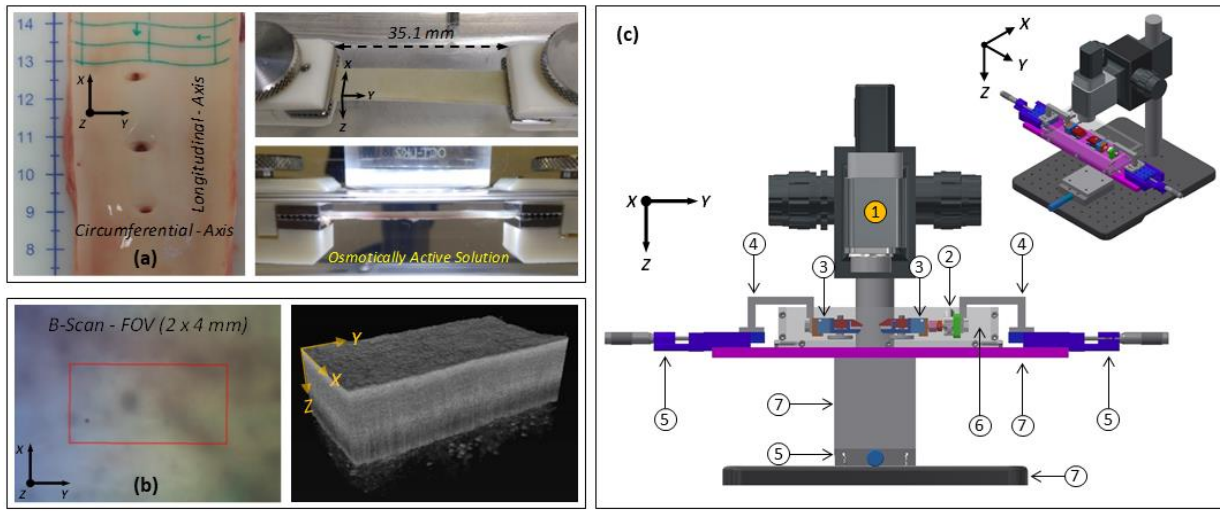

Figure 1.

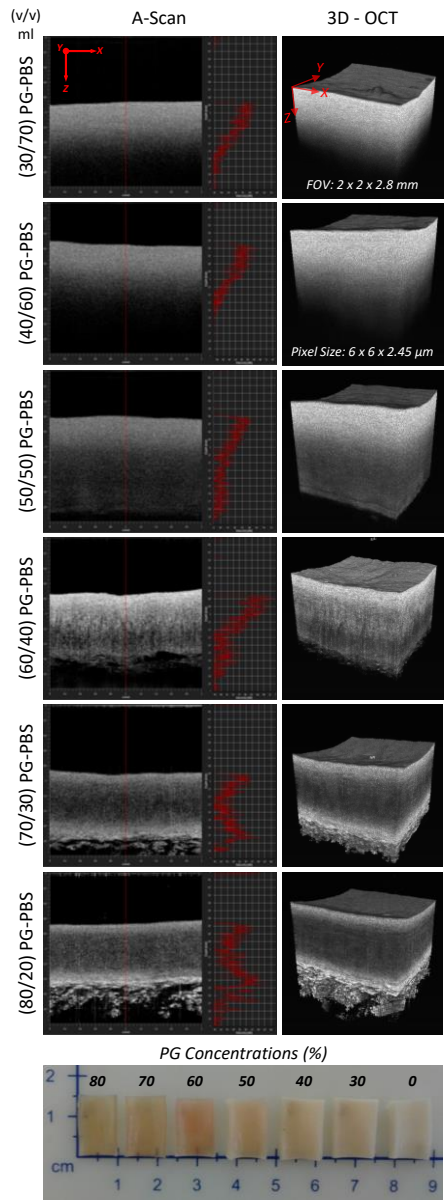

Figure 2.

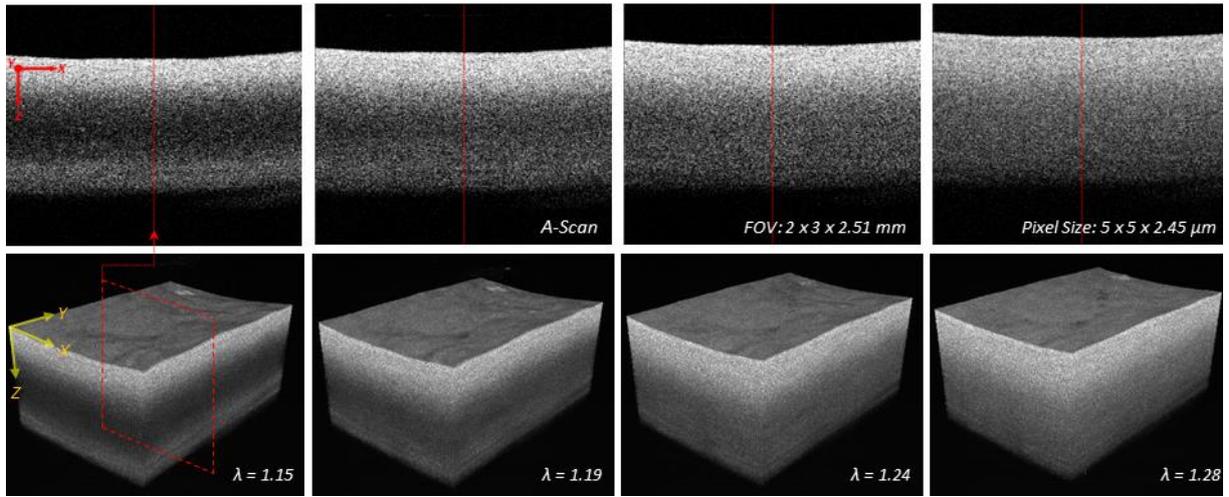

Figure 3.

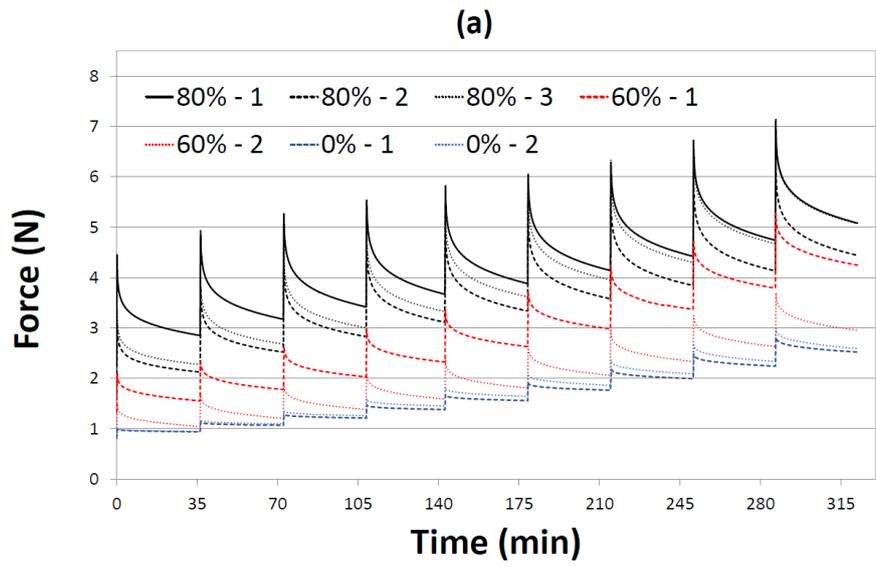
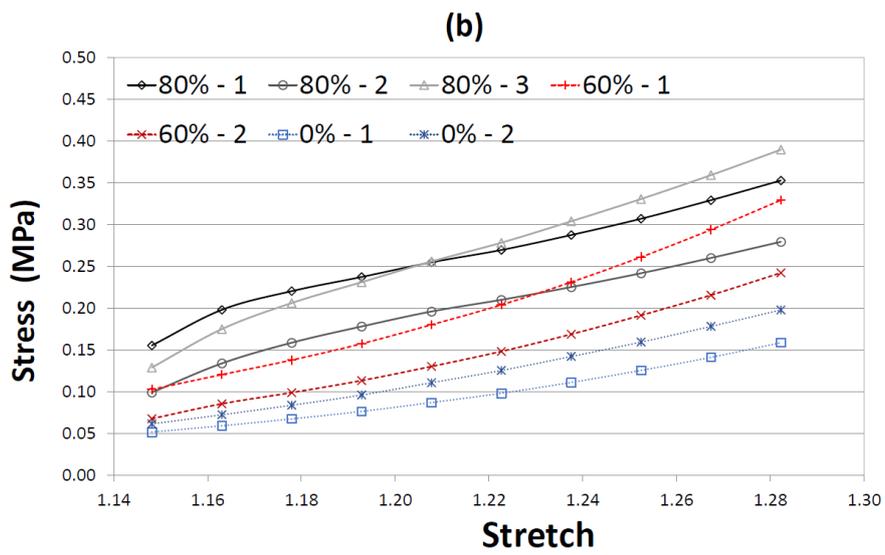

Figure 4.

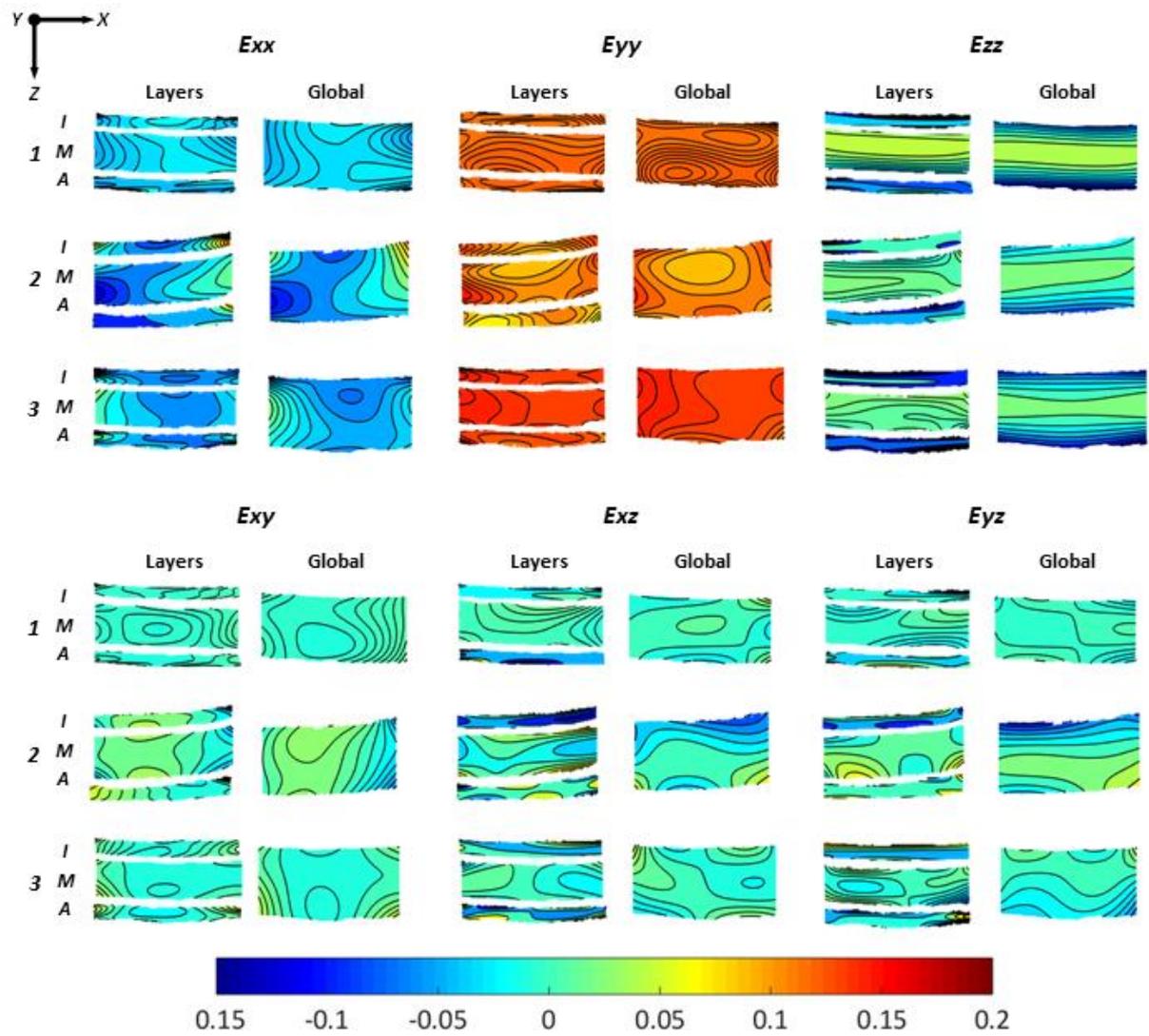

*Global, entire aortic wall; I, intima; M, media; A, adventitia.*

Figure 5.

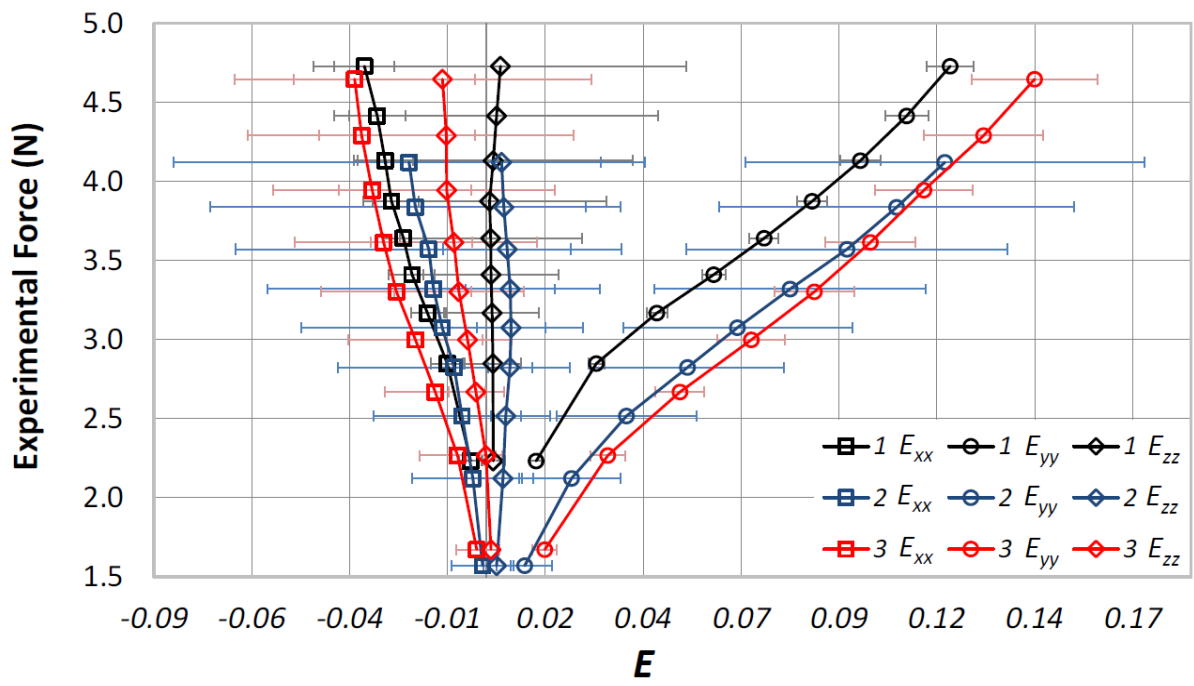

Figure 6.

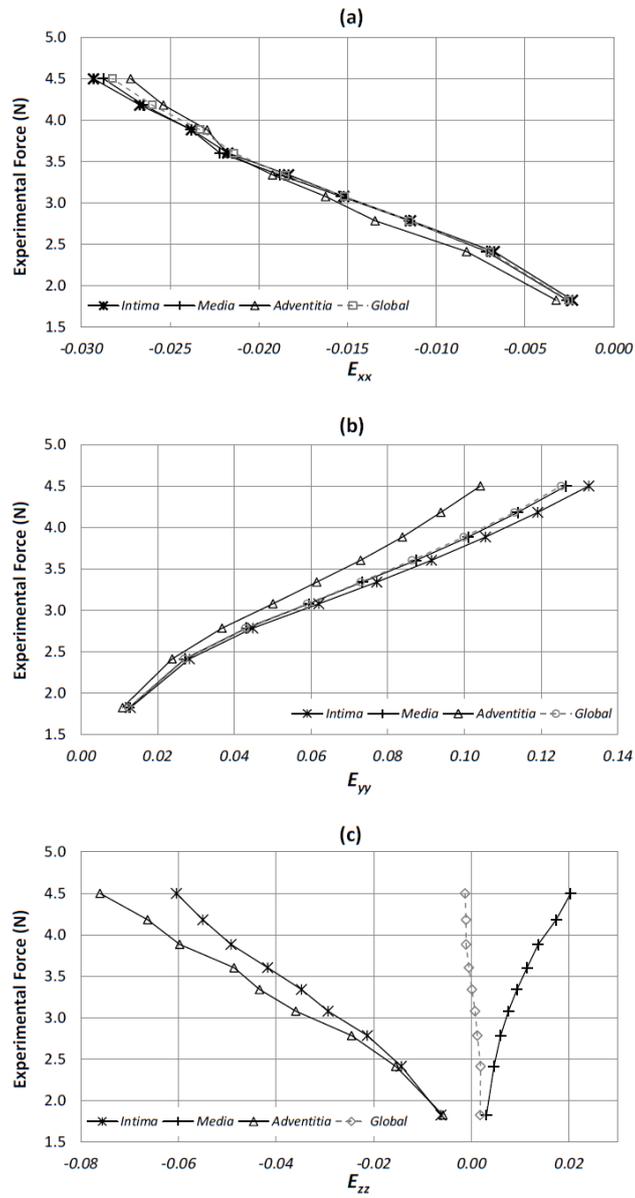

Figure 7.

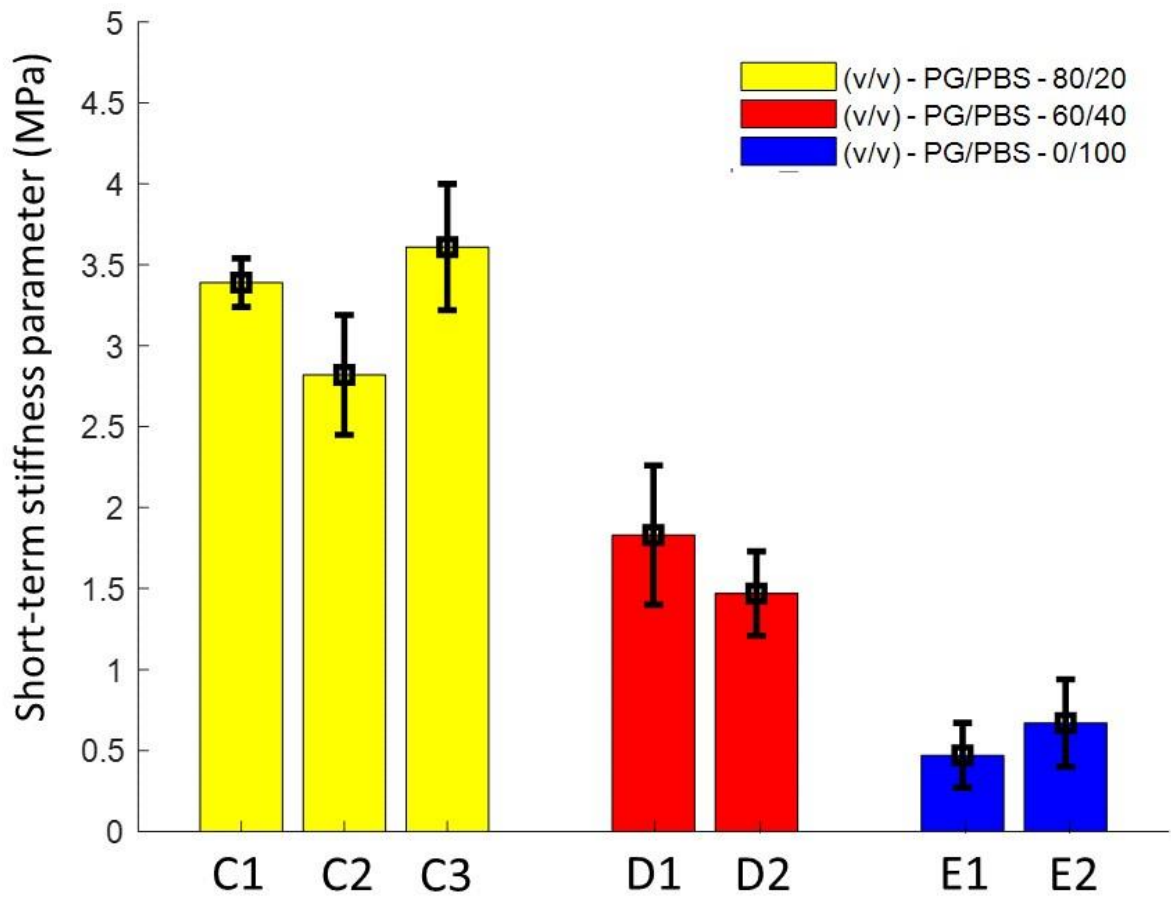

Figure 8.

## Appendix 1.

## Material modelling

To account for chemoelastic effects, the arterial tissue was modeled as a biphasic medium. The average Cauchy stress of the biphasic medium, denoted **T**, comprising the porous solid matrix and the interstitial fluid component, was written in a fully uncoupled Neo-Hookean form such as [9, 19]:

$$\mathbf{T} = -p\mathbf{I} + \kappa^s J^{-1}\ln(J)\mathbf{I} + 2\mu^s(\phi_0^f)J^{-1}\,\tilde{\mathbf{E}}^s \quad (A1)$$

where $p$ is the interstitial fluid pressure, $\kappa^s$ is the bulk modulus of the solid matrix, $\mu^s$ is the short-term stiffness parameter of the solid matrix (which depends on $\phi_0^f$, the initial matrix porosity – for the sake of simplifying the notation, the dependence on $\phi_0^f$ of $\mu^s$ will be omitted in the further equations), **I** is the identity tensor and $\tilde{\mathbf{E}}^s$ is a spatial strain tensor of the solid matrix satisfying

$$2\tilde{\mathbf{E}}^s = (\mathbf{B}^S - \mathbf{I}) - \frac{1}{3}\mathrm{Tr}(\mathbf{B}^S - \mathbf{I})\mathbf{I} = J^{-2/3}\left(\mathbf{B} - \frac{I_1}{3}\mathbf{I}\right) \quad (A2)$$

where **B** is the left Cauchy Green tensor of the mixture which satisfies $\mathbf{B}^S = J^{-2/3}\mathbf{B}$ where $\mathbf{B}^S$ is the left Cauchy Green tensor of the solid component, $I_1 = \mathrm{Tr}(\mathbf{B})$, and $J$ is the change of volume of the mixture (determinant of the deformation gradient **F** where $\mathbf{B} = \mathbf{F}\mathbf{F}^T$).

Combining Eqs (A1) and (A2), the deviatoric tensor of **T** can be written:

$$\mathrm{dev}(\mathbf{T}) = \mu^s J^{-5/3}\left(\mathbf{B} - \frac{I_1}{3}\mathbf{I}\right) \quad (A3)$$

Hyperelastic constitutive models of arteries commonly also include anisotropic supplemental terms which are functions of the stretch in preferred directions [49]. Although these terms could also be included to the biphasic model [9], they were not considered here as the fully uncoupled Neo-Hookean model could already suggest trends which were in agreement with the experimental results. Moreover, layer-specific constitutive properties were not considered either for the same reasons.

In uniaxial tension, assuming uniform stress and strains, it may be written:

$$\mathbf{F} = \begin{bmatrix} \lambda_x & 0 & 0 \\ 0 & \lambda_y & 0 \\ 0 & 0 & \lambda_z \end{bmatrix} \quad (A4)$$

$$\mathbf{T} = \begin{bmatrix} 0 & 0 & 0 \\ 0 & \sigma\lambda_y/J & 0 \\ 0 & 0 & 0 \end{bmatrix} \quad (A5)$$

where $\lambda_y$ is the stretch along the tensile axis $y$, $\lambda_x$ and $\lambda_z$ are the transverse stretches along $x$ and $z$, and $\sigma$ is the engineering stress or first Piola Kirchhoff stress ($\sigma = F/S_0$ where $F$ is the axial load and $S_0$ is the initial cross section area).

The deviatoric tensor of $\mathbf{T}$ can be written:

$$\mathrm{dev}(\mathbf{T}) = \sigma\lambda_y/(3J) \begin{bmatrix} -1 & 0 & 0 \\ 0 & 2 & 0 \\ 0 & 0 & -1 \end{bmatrix} \quad (A6)$$

Combining Eqs (A3) and (A6), it may be written:

$$\sigma = \mu^s J^{-2/3} \left( \lambda_y - \frac{\lambda_x^2}{2\lambda_y} - \frac{\lambda_z^2}{2\lambda_y} \right) \quad (A7)$$

In the following, we simplify $\lambda_y$ into $\lambda$, yielding:

$$\sigma = \mu^s J^{-2/3} \left( \lambda - \frac{\lambda_x^2 + \lambda_z^2}{2\lambda} \right) \quad (A8)$$

Differentiating Eq A8 with time, we may write [29, 50]:

$$\dot{\sigma} = \mu^s J^{-2/3} \left( 1 + \frac{\lambda_x^2 + \lambda_z^2}{2\lambda^2} \right) \dot{\lambda} - \mu^s J^{-2/3} \frac{\lambda_x \dot{\lambda}_x + \lambda_z \dot{\lambda}_z}{\lambda} - \frac{2}{3} \mu^s J^{-5/3} \left( \lambda - \frac{\lambda_x^2 + \lambda_y^2}{2\lambda} \right) \dot{J} \quad (A9)$$

where $\dot{\lambda}$, $\dot{\lambda}_x$, $\dot{\lambda}_z$ and $\dot{J}$ are the time derivatives of $\lambda$, $\lambda_x$, $\lambda_z$ and $J$ respectively.

Our loading protocol is decomposed into nine steps, and each step consists in a fast displacement ramp (phase 1) followed by a relaxation/swelling phase of about 30 min (phase 2).

According to [9], we can assume that fluid motion is negligible during the fast displacement ramp which represents the short term behavior ($\dot{J} = 0$ during phase 1).

Moreover, during phase 1, we assume that transverse strain rates are related to the axial strain rate according to the incompressibility assumption:

$$\frac{\dot{\lambda}_x}{\lambda_x} = \frac{\dot{\lambda}_z}{\lambda_z} = -\frac{\dot{\lambda}}{2\lambda} \qquad (A10)$$

Therefore, feeding Eq (A9) with Eq (A10), it may be written:

$$\dot{\sigma} = \mu^s J^{-2/3}\left(1 + \frac{\lambda_x^2 + \lambda_y^2}{\lambda^2}\right)\dot{\lambda} \qquad (A11)$$

During phase 2, the axial stress is maintained constant ($\dot{\lambda} = 0$) and we assume that the deformations are governed by fluid motions until reaching thermodynamical equilibrium at the end of phase 2. Therefore, Eq (A9) may be rewritten:

$$\dot{\sigma} = -\mu^s J^{-2/3}\frac{\lambda_x \dot{\lambda}_x + \lambda_z \dot{\lambda}_z}{\lambda} - \frac{2}{3}\mu^s J^{-5/3}\left(\lambda - \frac{\lambda_x^2 + \lambda_y^2}{2\lambda}\right)\dot{J} \qquad (A12)$$

Let us relate $\dot{\lambda}_x$, $\dot{\lambda}_z$ and $\dot{J}$ during phase 2:

$$\frac{\dot{J}}{J} = \frac{\dot{\lambda}_x}{\lambda_x} + \frac{\dot{\lambda}_z}{\lambda_z} \qquad (A13)$$

It is convenient to assume that, during phase 2, the strain rates in both transverse directions are proportional (they are not equal as shown by the experimental results)

$$\frac{\dot{\lambda}_x}{\lambda_x} = \gamma \frac{\dot{\lambda}_z}{\lambda_z} = \frac{\gamma}{\gamma+1}\frac{\dot{J}}{J} \qquad (A14)$$

where $\gamma$ is the proportionality factor relating the strain rates of the $x$ and $z$ directions.

Then Eq (A12) may be rewritten using Eqs (A13) and (A14):

$$\dot{\sigma} = -\mu^s J^{-5/3}\left(\frac{2}{3}\lambda + \frac{2\gamma-1}{3(\gamma+1)}\frac{\lambda_x^2}{\lambda} + \frac{2-\gamma}{3(\gamma+1)}\frac{\lambda_z^2}{\lambda}\right)\dot{J} \qquad (A15)$$

It was assumed that the pressure in the immersion bath is constant and arbitrarily set to zero. The fluid motion into the tissue is driven by the pressure drop between the immersion bath and the interior of the tissue. Let H denote the thickness of the permeable boundary layer over which this pressure drop is taking place at the beginning of phase 2, and let $A$ denote the area of the boundary. Then, Darcy's law may be used to define the flux of fluid entering the tissue such as:

$$Q = \dot{J}V_0 = -\frac{K(J)}{\mu^f}\frac{A}{H}(p + \pi(J)) \qquad (A16)$$

where $V_0$ is the volume of the tissue in the initial reference configuration, $\mu^f$ is the viscosity of the fluid, $K$ is hydraulic permeability of the tissue (which may depend on the degree of saturation or fraction of interstitial fluid), $\pi(J)$ is the difference of potential osmotic pressure between the immersion bath and the tissue (which decreases when the fraction of interstitial fluid due to the dilution effect [11, 18]) and $p$ is the hydrostatic pressure which can be derived from Eq (A1) according to:

$$p = -\text{Tr}(\mathbf{T})/3 + \kappa^s J^{-1}\ln(J) = -\sigma\lambda/(3J) + \kappa^s J^{-1}\ln(J) \quad (A17)$$

Then, Eq (A17) may be rewritten:

$$\dot{J} = -\frac{K(J)}{\mu^f}\frac{A}{HV_0}(-\sigma\lambda/(3J) + \kappa^s J^{-1}\ln(J) + \pi(J)) \quad (A18)$$

Finally, Eqs (A15) and (A18) make a system of ordinary differential equations in $J$ and $\sigma$. Then, $\sigma$ is subject to exponential decay as its decrease is an affine function of its current value when Eq A12 and A13 are combined:

$$\dot{\sigma} = -\frac{\mu^s K(J)}{\mu^f}\frac{A}{HV_0}\left(-\frac{\sigma\lambda}{3J} + \kappa^s J^{-1}\ln(J) + \pi(J)\right)J^{-5/3}\left(\frac{2}{3}\lambda + \frac{2\gamma-1}{3(\gamma+1)}\frac{\lambda_x^2}{\lambda} + \frac{2-\gamma}{3(\gamma+1)}\frac{\lambda_z^2}{\lambda}\right) \quad (A19)$$

The exponential decay constant is proportional to the stiffness $\mu^s$ of the solid matrix, to the hydraulic conductivity $K$ of the tissue and inversely to the viscosity of the interstitial fluid $\mu^f$.

The fluid motions actually give rise to heterogeneous stress and strain distributions across the tissue and the coupling between fluid motions and stress distributions require numerical resolutions to be solved. Finite element analyses of the biphasic model should be implemented to solve the whole transient response [9], whilst only the beginning of phase 2 can be approximated with Eq (A19) for the purpose of the current study. Another important improvement that should be considered to this model is anisotropy and layer-specificity of arteries [49].